\documentclass[floatfix, reprint, amsmath, amssymb, aps]{revtex4-2} 

\usepackage{graphicx, bbm}
\usepackage{dcolumn}
\usepackage{bm}
\usepackage[T1]{fontenc}
\usepackage{xfrac}

\newcommand{\RomanNumeralCaps}[1]
    {\MakeUppercase{\romannumeral #1}}

\begin{document}

\preprint{APS/123-QED}

\title{Revisiting the Problem of Positronium Moving Across a Magnetic Field}

\author{B.O. Kerbikov}
\email{bkerbikov@gmail.com}
\affiliation{%
 Lebedev Physical Institute, Moscow 119991, Russia
}%

\author{A.A. Simovonian}
\email{simovonian.aa@phystech.edu}
\affiliation{
 Moscow Institute of Physics and Technology,\\
 Dolgoprudny 141700, Moscow Region, Russia
}%


\begin{abstract}
Positronium spectrum and lifetimes are known with high precision. The situation is different for positronium moving across a magnetic field. The total momentum does not commute with the Hamiltonian and is replaced by the conserved pseudomomentum. The internal dynamics is not separated from the motion of the system as a whole. The Coulomb potential well is distorted and a wide outer potential well is created. We analytically determine the energy spectrum for a broad range of magnetic field and pseudomomentum values. The ground state energy is compared with the result obtained by solving the Bethe-Salpeter equation. The gauge independence of the pseudomomentum expectation value is established. On the pseudomomentum-magnetic field plane we locate the region in which the ground state resides in the outer well. The results may play a role in the suppression of pulsars' radio emission (polar cap problem). 
\end{abstract}

\maketitle


\section{Introduction}\label{sec:1}

There are several reasons for placing focus on theoretical and experimental studies of positronium (Ps). It is an ideal system for testing the accuracy of QED calculations with unprecedented precision \cite{1, 2, 3}. Ps may serve as a testing ground for the search for possible effects beyond the Standard Model \cite{1, 4, 5, 6, 7}. Currently, a vital interest in Ps stems from its possible role in suppressing one-photon $e^+e^-$ pair creation in neutron star (NS) magnetospheres \cite{8, 9, 10, 11}. Moving across a pulsar magnetic field (MF), which may be of the order of $10^{12}$ G, Ps undergoes a deep transformation. For sufficiently strong MF and adequate Ps velocity, an outer potential well is formed in addition to the Coulomb singularity. This wide and shallow well possesses an infinite discrete energy spectrum. The Ps constituents residing in it are separated by orders of magnitude more than they are in the Ps ground state. Accordingly, the annihilation rate exponentially drops with the damping constant proportional to the square of the Ps pseudomomentum (see below). The structure and properties of Ps moving in a strong MF are the subject of the present study.

For a system with no net electric charge, like the hydrogen atom or Ps, moving in MF, the center-of-mass (COM) momentum is not a conserved quantity. Complete separation of the internal dynamics from COM motion turns out to be impossible \cite{12, 13, 14, 15, 16, 17}. The spectrum and the wave function of the system parametrically depend on the pseudomomentum eigenvalues. Interaction between the COM and relative motion leads to the transformation of the spectrum and wave function. The Coulomb potential well (CW) is distorted, and above a certain value of the pseudomomentum an additional outer, aka magnetic, potential well (MW) is formed. For hydrogen such decentered states were predicted long ago in \cite{18} and have since been studied by a number of authors \cite{19, 20, 21, 22, 23}.
 
 To the best of the authors knowledge, for Ps the pseudomomentum $\textbf{K}$ was first introduced in \cite{24} and in a more elaborate form in \cite{25}. In these works calculations were performed for $\mathbf{K} = 0$ only. In \cite{26} the formation of Ps in the NS magnetosphere predicted in \cite{8} was reconsidered in the pseudomomentum formalism. In \cite{10} photon–Ps conversion in the pulsar magnetosphere was investigated with pseudomomentum implicitly introduced (see Sec.\ref{sec:7} below). It is important to note that in none of the above publications was the formation of the MW discussed, despite the fact that for the hydrogen atom this phenomenon has been known since 1976 \cite{18}. Delocalized states in Ps were first discussed in \cite{27, 28}. The authors performed a thorough study of the spectrum in the pseudomomentum formalism. Calculations in \cite{27, 28} were performed using the numerical adaptive finite elements method. By contrast, we rely on the analytical methods different for various intervals of MF strength and K values. Explicit formulas derived in the present work allow to analyze the dependence of the spectrum on $\text{B}$ and $\text{K}$ values and to investigate asymptotic regimes. Some overlap between the present research and \cite{27, 28} is unavoidable.

It is necessary to remind some basic Ps and MF properties and characteristics. Throughout this work we follow the conditions $\hbar = c = 1$, $\alpha = e^2 = 1/137$. The ground state binding energy of Ps is $\text{E}_{\text{B}} = - m e^4/4 \simeq -6.8 \text{ eV} = -R_{\infty}/2$, where $m \equiv m_e = 0.51 \text{MeV}$. Ps Bohr radius is $a_{\text{B}}(\text{Ps}) = 2/me^2 = 2 a_0 \simeq 1.06\cdot10^{-8} \text{ cm}$, where $a_0 \simeq 1/m e^2 = 0.529 \cdot 10^{-8}$ cm is the hydrogen Bohr radius. In the above system of units $1 \text{ MeV}^2 = 1.45 \cdot 10^{13} \text{ G}$. MF Landau radius is $\ell = 1/\sqrt{e\text{B}}$, the cyclotron frequency is $\omega_c = e \text{B}/m$, $\text{B}_a = m^2 e^3 = 2.35 \cdot 10^{9} \text{ G}$ is the atomic unit of MF strength, $\ell(\text{B}_a) = a_0$. The atomic field strength $\text{B}_a'$ for Ps defined as $l = 1/\sqrt{e \text{B}_a'} = a_{\text{B}}(\text{Ps})$ is equal to $\text{B}_a' = \text{B}_a/4$. Our main focus will be on MF in the range $\gamma = \text{B}/\text{B}_a > 1$.

The work is organized as follows. In Sec.\ref{sec:2} we formulate the Hamiltonian of two particles with opposite electric charges moving in MF. The pseudomomentum $\mathbf{K}$ is introduced and it is shown that the internal wave function and the energy spectrum depend on the value of K. In Sec.\ref{sec:3} the critical value $\text{K}_c$ for the formation of the MW is derived. The energy spectrum equation in the adiabatic approximation and in the «shifted» representation is derived in Sec.\ref{sec:4}. In Sec.\ref{sec:5} the configuration with two separated potential wells is investigated. In Sec.\ref{sec:6} the situation when the potential wells overlap is considered. In Sec.\ref{sec:7} we consider issue of the gauge choice in the Ps problem. Sec.\ref{sec:8} contains the comparative discussion of the Ps problem solution in different gauges and approximations. In Sec.\ref{sec:9} we consider some physical implications of the results. It includes the emergence of long-lived Ps, giant dipole moment and the problem of one-photon Ps annihilation. Sec.\ref{sec:10} contains the summary of the work.

\section{Two coupled particles with opposite charges moving across magnetic field}\label{sec:2}

The physics of bound quantum system moving in magnetic field (MF) is intricate. Due to the lack of translational invariance, the total kinetic (mechanical) momentum $\hat{\mathbf{P}}_{\text{kin}} = \sum_{i = 1}^N (\hat{\mathbf{p}}_i - e_i \mathbf{A}_i)$ is not a conserved quantity \cite{12, 13, 14, 15, 16, 17, 18, 19, 20, 21, 22, 23}. For a neutral system like Ps, it is possible to construct a quantity $\mathbf{K}$, called the pseudomomentum, which commutes with the Hamiltonian \cite{13, 14, 15, 16, 17}. The Hamiltonian of two particles with opposite electric charges $e_1 = e > 0$, $e_2 = -e$ and equal masses $m_1 = m_2 = m$ in a constant MF can be written as
\begin{equation}\label{eq:Hamiltoniantwoparticles}
	\hat{\mathcal{H}} = \frac{1}{2 m} \bigg[ \hat{\mathbf{p}}_1-e \mathbf{A}(\mathbf{r}_1)\bigg]^2 + \frac{1}{2 m} \bigg[ \hat{\mathbf{p}}_2 + e \mathbf{A}(\mathbf{r}_2) \bigg]^2 + V(\boldsymbol{{\eta}}),
\end{equation}
where $\boldsymbol{\eta} = \mathbf{r}_1 - \mathbf{r}_2$, $V(\boldsymbol{\eta}) = - e^2/|\boldsymbol{\eta}|$ for Ps. Spin-dependent and $\boldsymbol{\sigma} \mathbf{B}$ terms are temporarily omitted. MF is assumed to be homogeneous and directed parallel to the $z$-axis: $\mathbf{B} = (0, 0, \text{B})$. We use the symmetric gauge $\mathbf{A} = \sfrac{1}{2}~ \mathbf{B} \times \mathbf{r} = \sfrac{1}{2} ~ \text{B}(-y, x, 0)$. Next we introduce the center-of-mass coordinate $\mathbf{R} = \sfrac{1}{2}~(\mathbf{r}_1 + \mathbf{r}_2)$ and momentum operators
\begin{equation}
	\hat{\boldsymbol{\pi}} = -i \frac{\partial}{\partial \mathbf{\boldsymbol{\eta}}}, ~~~~~ \hat{\mathbf{P}} = \hat{\mathbf{p}}_1 + \hat{\mathbf{p}}_2 = -i \frac{\partial}{\partial \mathbf{R}}.
\end{equation}

The Hamilton operator takes the form
\begin{eqnarray}\label{eq:HaminReta}
	\hat{\mathcal{H}} &=& \frac{1}{4 m} \bigg( -i \frac{\partial}{\partial \mathbf{R}} - \frac{e}{2} \mathbf{B} \times \boldsymbol{\eta}\bigg)^2 + \nonumber\\& & + \frac{1}{m} \bigg( \boldsymbol{\pi} - \frac{e}{2} \mathbf{B}\times \mathbf{R}\bigg)^2 + V(\boldsymbol{\eta})~.
\end{eqnarray}

The pseudomomentum operator commuting with $\hat{\mathcal{H}}$ reads 
\begin{equation}\label{eq:pseudomomentum}
	\hat{\mathbf{K}} = \sum\limits_{i = 1}^{2}\bigg( \mathbf{p}_i + \frac{1}{2} e_i \mathbf{B} \times \mathbf{r}_i \bigg) = - i \frac{\partial}{\partial \mathbf{R}} + \frac{e}{2} \mathbf{B} \times \boldsymbol{\eta}.
\end{equation}

$\hat{\mathbf{K}}$ is the integral of motion and eigenfunctions $\Psi(\mathbf{R}, \boldsymbol{\eta})$ of $\hat{\mathcal{H}}$ are eigenfunctions of $\hat{\mathbf{K}}$
\begin{equation}\label{eq:spetrumpseudomomentum}
	\hat{\mathbf{K}} \Psi(\mathbf{R}, \boldsymbol{\eta}) = \mathbf{K} \Psi(\mathbf{R}, \boldsymbol{\eta}),
\end{equation}
where $\mathbf{K}$ is the eigenvalue of $\hat{\mathbf{K}}$. Let us represent $\Psi(\mathbf{R}, \boldsymbol{\eta})$ as $\Psi(\mathbf{R}, \boldsymbol{\eta}) = \exp(i\boldsymbol{\nu} \mathbf{R}) \varphi_{\mathbf{K}}(\boldsymbol{\eta})$ with yet unknown $\boldsymbol{\nu}$. It can be found from (\ref{eq:pseudomomentum}) and (\ref{eq:spetrumpseudomomentum}) that
\begin{eqnarray}
\hat{\mathbf{K}} \Psi(\mathbf{R}, \boldsymbol{\eta}) = \bigg(\boldsymbol{\nu} + \frac{e}{2} \mathbf{B} \times \boldsymbol{\eta} \bigg) \exp(i \boldsymbol{\nu} \mathbf{R}) \varphi_{\mathbf{K}}(\boldsymbol{\eta}) & = & \nonumber \\ = \mathbf{K} \exp(i \boldsymbol{\nu} \mathbf{R}) \varphi_{\mathbf{K}}(\boldsymbol{\eta}),
\end{eqnarray}
hence $\mathbf{K} = \boldsymbol{\nu} + (e/2) \mathbf{B} \times \boldsymbol{\eta}$, and we get $\Psi(\mathbf{R}, \boldsymbol{\eta})$ in the form
\begin{equation}\label{eq:wavefunction}
	\Psi(\mathbf{R}, \boldsymbol{\eta}) = \exp\bigg[i \bigg(\mathbf{K} - \frac{e}{2} \mathbf{B} \times \boldsymbol{\eta}\bigg) \mathbf{R} \bigg] \varphi_{\mathbf{K}}(\boldsymbol{\eta}).
\end{equation}

The action of $\hat{\mathcal{H}}$ given by (\ref{eq:HaminReta}) on the factorized wave function (\ref{eq:wavefunction}) yields the eigenvalue equation for $\varphi_{\mathbf{K}}(\boldsymbol{\eta})$
\begin{eqnarray}\label{eq:ShEqpseu}
	\bigg[ \frac{\hat{\boldsymbol{\pi}}^2}{m} + \frac{\mathbf{K}^2}{4 m} - \frac{e}{2 m} [\mathbf{K} \times \mathbf{B}] \cdot \boldsymbol{\eta} + \frac{e^2}{4 m}[\mathbf{B}\times \boldsymbol{\eta}]^2 + & \nonumber\\ + V(\boldsymbol{\eta})\bigg] \varphi_{\mathbf{K}}(\boldsymbol{\eta}) = \text{E}\cdot \varphi_{\mathbf{K}}(\boldsymbol{\eta}).
\end{eqnarray}

We see that unlike the free-field case, the internal wave function $\varphi_{\mathbf{K}}(\boldsymbol{\eta})$ as well as the eigenvalue $\text{E}$ depend on $\mathbf{K}$. Collective and internal motion are connected through the motional Stark term $e/2 m[\mathbf{K}\times\mathbf{B}] \cdot \boldsymbol{\eta}$. The electric field induced by this term is directed perpendicular to $\mathbf{B}$. It is instructive to rewrite $\hat{\mathcal{H}}$ in terms of the pseudomomentum. This allows one to relate the pseudomomentum to the Ps center-of-mass motion kinetic energy and velocity. From (\ref{eq:Hamiltoniantwoparticles}) and (\ref{eq:pseudomomentum}) one easily gets
\begin{eqnarray}\label{eq:hampseu}
	\hat{\mathcal{H}} & = & \frac{1}{4 m} \bigg( \hat{\mathbf{K}} - e \mathbf{B} \times \boldsymbol{\eta}\bigg)^2 - \nonumber \\ & & - \frac{1}{m} \bigg(\boldsymbol{\pi} - \frac{e}{2} \mathbf{B}\times \mathbf{R}\bigg)^2 + V(\boldsymbol{\eta}).
\end{eqnarray}

The mechanical momentum of Ps reads 
\begin{equation}
    \mathbf{P} = \sum\limits_{j}\bigg(- \frac{\partial}{\partial\mathbf{r}_j} - e_j \mathbf{A}_j\bigg) = - i \frac{\partial}{\partial \mathbf{R}} - \frac{1}{2} e \mathbf{B}\times\boldsymbol{\eta}.
\end{equation}
The derivative $\partial/\partial \mathbf{R}$ and $\hat{\mathbf{K}}$ are related according to (\ref{eq:pseudomomentum}). This gives the desired result 
\begin{equation}
    \mathbf{P} \Psi(\mathbf{R}, \boldsymbol{\eta}) = (\hat{\mathbf{K}} - e \mathbf{B}\times \boldsymbol{\eta})\Psi(\mathbf{R}, \boldsymbol{\eta}).
\end{equation}
This makes it clear that the first term in (\ref{eq:hampseu}) in the center-of-mass kinetic energy.

Another way to identify the first term in (\ref{eq:hampseu}) with the Ps kinetic energy is to use the Hamilton's equation
\begin{equation}\label{eq:velocity}
	\mathbf{V} = \dot{\mathbf{R}} = \frac{\partial \mathcal{H}}{\partial \mathbf{K}} = \frac{1}{2 m} \bigg( \mathbf{K} - e \mathbf{B} \times \boldsymbol{\eta} \bigg).
\end{equation}

For a given MF the velocity $\mathbf{V}$ of Ps is determined by pseudomomentum and the electric dipole moment. The expectation value of $\mathbf{V}$ is obtained upon averaging (\ref{eq:velocity}) over Ps wave function.

\section{The Outer Potential Well Formation}\label{sec:3}

The motion of Ps across a magnetic field (MF) results in a transformation of the potential shape. For a fixed value of MF a second potential well begins to form at a certain critical value $\text{K}_c$ of the pseudomomentum. This phenomenon has been studied for the hydrogen atom \cite{18, 19, 20, 21, 22, 23}, Ps \cite{27, 28}, and quarkonium \cite{29}.

From (\ref{eq:ShEqpseu}) and (\ref{eq:hampseu}) we may write the internal motion effective potential $\text{U}_{\text{eff}} (\boldsymbol{\eta})$ as 
\begin{eqnarray}\label{eq:effpot}
	\text{U}_{\text{eff}}(\boldsymbol{\eta}) &=& \frac{\mathbf{K}^2}{4 m} + \frac{e \text{B}}{2 m} \text{K}_x y - \frac{e \text{B}}{2 m} \text{K}_y x + \nonumber\\& & + \frac{e^2 \text{B}^2}{4 m} (x^2 + y^2) - \frac{e^2}{\sqrt{x^2 + y^2 + z^2}},
\end{eqnarray}
where $\boldsymbol{\eta} = (x, y, z)$. The first term is spatially independent and acts as an additive constant to the binding energies. The evolution of the potential shape as a function of $\text{K}$ is governed mainly by the interplay between the Stark $x, y$ terms and the diamagnetic fourth term in (\ref{eq:effpot}). The potential possesses azimuthal symmetry, and one can set either $\text{K}_x$ or $\text{K}_y$ equal to zero. We choose $K_y = 0$, so that $\mathbf{K} = (\text{K}, 0, 0)$. The projection of (\ref{eq:effpot}) onto the $x = z = 0$ plane is
\begin{equation}\label{eq:effpot1}
	\text{U}_{\text{eff}} (y) = \frac{\mathbf{K}^2}{4 m} + \frac{e \text{B}}{2 m} \text{K}y + \frac{e^2 \text{B}^2}{4 m}y^2 - \frac{e^2}{|y|}.
\end{equation}

 The condition for the potential minimum at $y < 0$ $\partial \text{U}_{\text{eff}} / \partial y = 0$ yields the equation 
\begin{equation}\label{eq:cubiceq}
	y^3 + \frac{\text{K}}{e \text{B}} y^2 - \frac{2 m}{\text{B}^2} = 0.
\end{equation}
 
For the outer magnetic well (MW) and a saddle point (SP) to exist for $\text{K} > 0, y < 0$ this equation must have three real roots. It implies the following condition on $\text{K}$ and $\text{B}$
\begin{equation}\label{critpseudomomentum}
	\text{K}^3 > \text{K}_c^3 = \frac{27}{2} (m e^2) (e \text{B}) = \frac{27}{2} \bigg( \frac{e \text{B}}{a_0} \bigg).
\end{equation}

The three roots are
\begin{eqnarray}\label{eq:roots1}
	y_1 & = & \frac{\text{K} \ell^2}{3} \bigg[ 2 \cos \frac{\alpha}{3} - 1\bigg], \\ & & y_{2, 3} = -\frac{\text{K} \ell^2}{3} \bigg[1 + \cos \frac{\alpha}{3} \pm \sqrt{3} \sin \frac{\alpha}{3} \bigg], \label{eq:roots2}
\end{eqnarray}
where 
\begin{equation}
    \cos \alpha = \frac{27}{a_0 \ell^2 \text{K}^3} - 1 = \frac{2 \text{K}_c^3}{\text{K}^3} - 1.
\end{equation}
 Therefore $y_2 < y_3 < 0 < y_1$. It means that $y_1$ is an unphysical solution, $y_2 \equiv y_m$ corresponds to the MW minimum, $y_3 \equiv y_s$ -- to the saddle point. Note that at fixed value of MF and increasing $\text{K} \gg \text{K}_c$ (\ref{eq:roots2}) yields
 \begin{eqnarray}\label{eq:roots}
 	y_m \rightarrow - \frac{\text{K}}{e \text{B}} &=& - \text{K} l^2, y_s \rightarrow - \sqrt{\frac{2em}{\text{B}\text{K}}} = - l\sqrt{\frac{2}{a_0 \text{K}}}, \nonumber\\ & & \frac{y_m}{y_s} \rightarrow \frac{3\sqrt{3}}{2}\bigg( \frac{\text{K}}{\text{K}_c} \bigg)^{3/2}.
 \end{eqnarray}
This shows that with increasing $\text{K}$, the MW tends to separate from CW for any given value of the MF. In the same limit, $\text{K} \gg \text{K}_c$, the MW minimum approaches zero energy, from below: $\text{U}_{\text{eff}}(y_m) \rightarrow - e^2 (e \text{B})/\text{K}$.

\begin{figure}
\center{\includegraphics[width = 1.0 \linewidth, keepaspectratio]{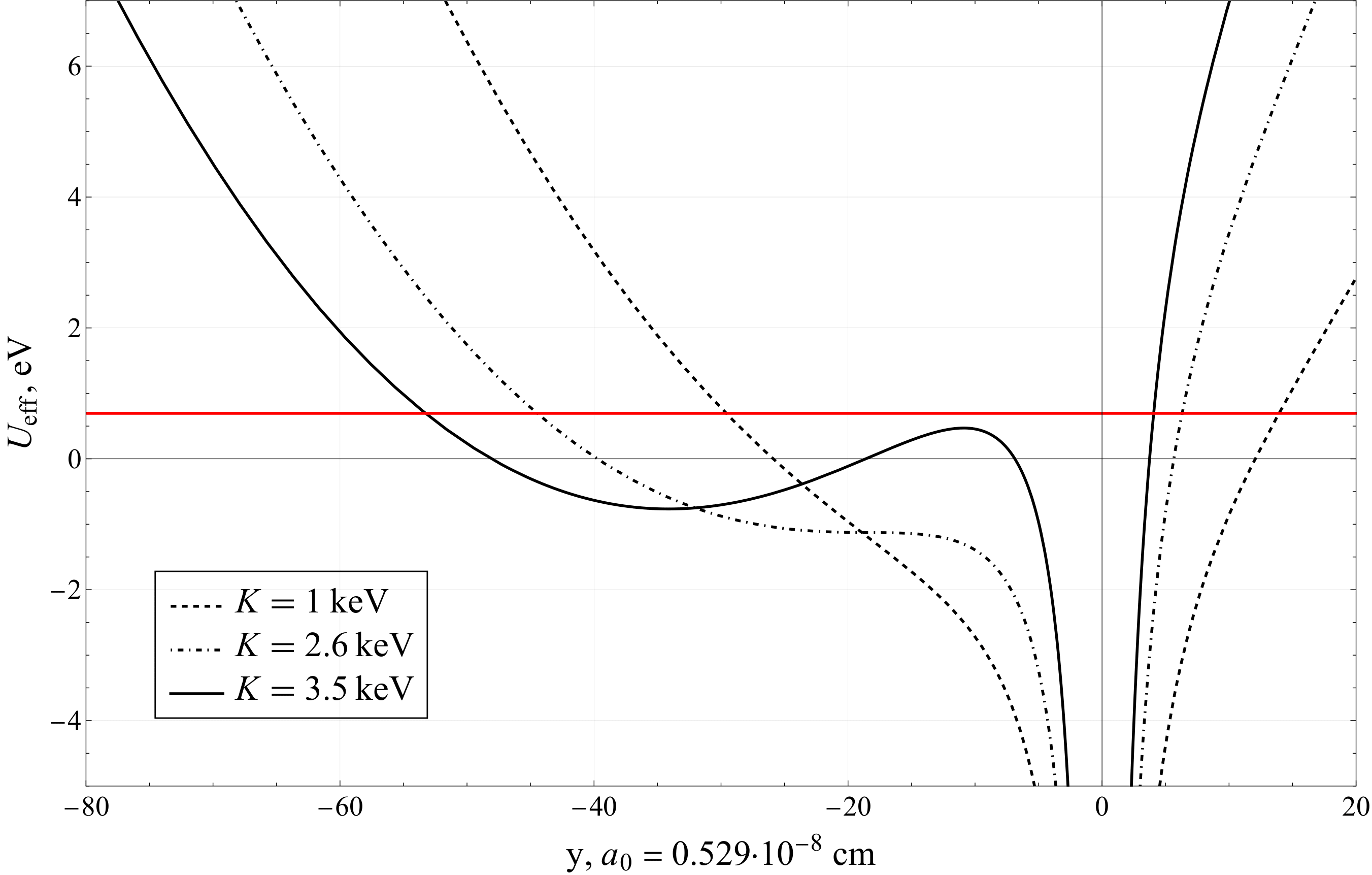}}
\caption{The effective potential (\ref{eq:effpot1}) in eV as a function of y in $a_0 = 1/m_e e^2 = 0.529 \cdot 10^{-8}$ cm for MF $\text{B} = 6 \cdot 10^7$ G, $\text{K}_c = 2.61$ keV and three different values of $\text{K}$. The solid red horizontal line corresponds to the ionization threshold $\text{I} = e \text{B}/m$.}\label{fig:1}
\end{figure}

The effective potential (\ref{eq:effpot1}) along the $y$ direction is shown in Fig.~\ref{fig:1} for $\text{B} = 6 \cdot 10^7$ G and three different values of $\text{K}$. The solid red horizontal line corresponds to $\text{I} = e\text{B}/m$. The quantity $\text{I}$ is the ionization potential. In the absence of the Coulomb interaction, it is equal to the sum of the individual electron and positron Landau ground state energies.

The motion in a plane perpendicular to the direction of the MF is at large distances dominated by the contribution of the diamagnetic term. This term provides confinement and ionization is possible only along the MF direction. The presence of the cubic term in (\ref{eq:cubiceq}) makes the evolution of the effective potential with $\text{K}$, as plotted in Fig.~\ref{fig:1}, reminiscent of a first-order phase transition.

\section{The Energy Spectrum Equation}\label{sec:4}

The spectral problem (\ref{eq:ShEqpseu}) for the Hamiltonian 
\begin{eqnarray}
	\hat{\mathcal{H}} = \frac{\mathbf{K}^2}{4m} - \frac{1}{m}\Delta_{\boldsymbol{\eta}} &-& \frac{e}{2m}[\mathbf{K}\times \mathbf{B}]\cdot \boldsymbol{\eta} + \nonumber\\ &+&  \frac{e^2}{4m}[\mathbf{B}\times \boldsymbol{\eta}]^2 - \frac{e^2}{|\boldsymbol{\eta}|} \label{eq:Hamw/os}
\end{eqnarray}
does not admit an exact solution. For the exciton and the hydrogen atom moving in a MF, some analytical approximations have been developed \cite{13, 16, 18, 19, 20}. Numerical calculations for Ps have been performed in \cite{27, 28}. We aim to solve the problem relying on analytical methods as far as possible. Following \cite{13, 20, 22}, it is appropriate to use the «shifted» representation $\boldsymbol{\eta}' = \boldsymbol{\eta} - \boldsymbol{\eta}_c$, where $\boldsymbol{\eta}_c$ is the difference between the coordinates of the gyro-motion guiding centers of $e^+$ and $e^-$
\begin{eqnarray}\label{eq:gyromotion}
	\boldsymbol{\eta}_c &=& \frac{\mathbf{K}_1\times \mathbf{B}}{e \text{B}^2} - \frac{\mathbf{K}_2\times \mathbf{B}}{-e \text{B}^2} = \nonumber \\ & & = - \frac{[\mathbf{B} \times \mathbf{K}]}{e \text{B}^2} = - \frac{2m}{e \text{B}^2}[\mathbf{B} \times \mathbf{V}],
\end{eqnarray}
where $\mathbf{V}$ is given by (\ref{eq:velocity}). Note that with $\mathbf{B}$ directed along the $z$-axis $\eta_{cz} = 0$. Henceforth we shall use the following notations: $\boldsymbol{\eta}' = (\eta'_x, \eta'_y, z) \equiv (\boldsymbol{\rho}', z)$, $\boldsymbol{\eta}_c = (\eta_{cx}, \eta_{cy}) \equiv \boldsymbol{\rho}_0$ and the coordinate $z$ is not affected by the shift. Inserting $\boldsymbol{\eta}' = \boldsymbol{\eta} - \boldsymbol{\eta}_c$ into (\ref{eq:Hamw/os}) one obtains the «shifted» Hamiltonian
\begin{eqnarray}\label{eq:Hamw/s}
	\hat{\mathcal{H}}' &=& \frac{\text{K}_z^2}{4 m} - \frac{1}{m}\frac{\partial^2}{\partial z^2} - \frac{1}{m}\bigg( \frac{\partial^2}{\partial \eta_x'^2} + \frac{\partial^2}{\partial \eta_y'^2} \bigg) + \nonumber \\ & & +  \frac{e^2}{4 m} \mathbf{B}^2 \boldsymbol{\rho}'^2 - \frac{e^2}{\sqrt{(\boldsymbol{\rho}' + \boldsymbol{\rho}_0)^2 + z^2}}.
\end{eqnarray}
The Stark term has been eliminated in this representation. The pseudomomentum $\mathbf{K}$ enters into (\ref{eq:Hamw/s}) through $\boldsymbol{\rho}_0$ and only then via an explicit term $\text{K}^2_z/4m$.

It is well known that the Schrodinger equation for the hydrogenlike system (hydrogen atom, Ps, exciton) under the presence of a uniform MF does not admit exact analytical solution \cite{30}. The Hamiltonians (\ref{eq:Hamw/os}) and (\ref{eq:Hamw/s}) are of this type. However, in strong MF when $\gamma = \text{B}/\text{B}_a \gg 1$, $a_0 \gg \ell$ the Coulomb potential can be considered as a perturbation in the background MF. Then the Schrodinger equation can be solved using the adiabatic approximation \cite{31, 32}. The wave unction is expanded over the complete set of Landau orbitals \cite{30}. Only the term corresponding to the lowest Landau (LLL) level is retained in this expansion. For the Hamiltonian (\ref{eq:Hamw/s}) the adiabatic approximation takes the form
    \begin{equation}\label{eq:phi}
        \varphi_{\mathbf{K}}(\boldsymbol{\rho}', z) = R(\boldsymbol{\rho}') f(z),
    \end{equation}
where $R(\boldsymbol{\rho}')$ is the LLL wave function
\begin{equation}\label{eq:R}
    R(\boldsymbol{\rho}') = \frac{1}{\sqrt{2 \pi}\ell} \exp\bigg( - \frac{\boldsymbol{\rho}'^2}{4 \ell^2} \bigg),
\end{equation}
and $f(z)$ is a longitudinal part. Adiabatic approximation is increasingly accurate with $\gamma$ increasing and for $\gamma \rightarrow \infty$ it becomes exact. The real upper bound is $\gamma \sim 10^4$ corresponding to the critical, or Schwinger field $\text{B}_{\text{cr}} = m^2/e = 4.414 \cdot 10^3$ G for which the energy distance in the Landau spectrum is equal to $m_e$ and relativistic effects become non-negligible. There is no a universal formula expressing the accuracy of the adiabatic approximation as a function of $\gamma$. To estimate the error one can evaluate perturbatively the matrix element of the Coulomb potential between the LLL and the next level \cite{37}. The common way to verify the accuracy is the comparison with numerical calculations \cite{38}. Such a program has been fulfilled during the past decades for the non-moving hydrogen atom in MF, see \cite{22, 38, 1044} and references therein. For the astrophysical applications it is important that for $\text{B} \sim 10^{12}$ G the accuracy is high. For moving hydrogen atom the adiabatic approximation has been considered in \cite{18, 1045}.

We return now to Ps problem under consideration. Substituting (\ref{eq:phi})--(\ref{eq:R}) into the Schrödinger equation $\hat{\mathcal{H}}' \varphi_{\mathbf{K}} = \text{E} \varphi_{\mathbf{K}}$, acting by $\partial^2/\partial \boldsymbol{\rho}'^2$ on $R(\boldsymbol{\rho}')$, multiplying by $R(\boldsymbol{\rho}')$ and integrating over $d\boldsymbol{\rho}'$, we obtain
\begin{eqnarray}
	\label{eq:forz}
	\frac{d^2 f}{dz^2} + m \bigg[\text{E} - \frac{e \text{B}}{m} + \text{U}(z)\bigg] f(z) = 0,  \\
	\text{U}(z) = \frac{e^2}{2 \pi \ell^2} \iint\limits_{\mathbb{R}^2} d\boldsymbol{\rho}' \frac{\exp(- \boldsymbol{\rho}'^2/2\ell^2)}{\sqrt{(\boldsymbol{\rho}' + \boldsymbol{\rho}_0)^2 + z^2}}. \label{eq:potforz1}
\end{eqnarray}
The term $\text{K}^2_z/4m$ in (\ref{eq:Hamw/s}) does not influence the spectrum and is omitted.

Transition from the Schrödinger equation $\mathcal{H}'\varphi_{\mathbf{K}} = \text{E}\varphi_{\mathbf{K}}$ with $\mathcal{H}'$ given by (\ref{eq:Hamw/s}) to (\ref{eq:forz}) may be considered as averaging over the fast MF variables.

As already noted, $\text{I} = e\text{B}/m$ in (\ref{eq:forz}) is zero-point energy of the LLL, or the ionization threshold. The energy $\text{E}$ in (\ref{eq:forz}) is given by
\begin{equation}\label{eq:defeb}
    \text{E} = \text{I} + \text{E}_{\text{B}},
\end{equation}
where $\text{E}_{\text{B}}$ is the binding energy. The sates with $(\text{E}-\text{I}) < 0$ are bound states corresponding to the closed channels. States with $(\text{E}-\text{I}) > 0$ lie above the ionization threshold and form a series of autoionizing resonances. With this definition (\ref{eq:forz}) takes the form 
\begin{equation}\label{eq:longforeb}
    \bigg( - \frac{1}{m}\frac{d^2}{dz^2} - \text{U}(z) \bigg) f(z) = \text{E}_{\text{B}} f(z).
\end{equation}

\section{The Separated Potential Wells}\label{sec:5}

According to (\ref{critpseudomomentum}) the outer potential well is formed when $\text{K}^3 > \text{K}_c^3 = (27/2) m e^3 \text{B} = (27/2)(1/a_0 \ell^2)$. Under this condition different configurations of the two potential wells and a barrier between them are possible. The shape of the resulting configuration dictates the most appropriate methods to solve eq.(\ref{eq:forz}). The problem has three parameters with the dimension of length
\begin{equation}\label{eq:sep_pot_wells}
	a_{\text{B}}(\text{Ps}) = 2 a_0 = \frac{2}{m e^2}, \ell = \frac{1}{\sqrt{e \text{B}}}, |\boldsymbol{\rho}_0| = |y_0| = \frac{\text{K}}{e \text{B}}.
\end{equation}
Here $|y_0|$ is the absolute value of the projection of $\boldsymbol{\rho}_0 = - [\mathbf{B}\times \mathbf{K}]/e\text{B}^2$ onto the $x=z=0$ plane and $\mathbf{K} = (\text{K}, 0, 0)$ in this projection. Important to note that according to (\ref{eq:roots}) $y_0 = - \text{K}/e\text{B}$ coincides with the position of the MW minimum $y_m$ in the limit $\text{K} \gg \text{K}_c$. Therefore one can interpret $|y_0|$ as the distance from the bottom of the MW to the Coulomb center. The two wells, the Coulomb and the magnetic ones, are actually separated if $|y_0| \gg a_{\text{B}}(\text{Ps})$. The use of the adiabatic approximation (\ref{eq:phi}) implies that $a_{\text{B}}(\text{Ps}) \gg \ell$. We come to the conclusion that the conditions for the double-well regime in the adiabatic approximation are the following
\begin{equation}\label{eq:inequal}
	\text{K} > \text{K}_c, ~~~~~ |y_0| \gg a_{\text{B}}(\text{Ps}) \gg \ell.
\end{equation}
For the hydrogen atom the above situation was investigated in \cite{18}.
	
\begin{figure}
\center{\includegraphics[width = 1.0 \linewidth]{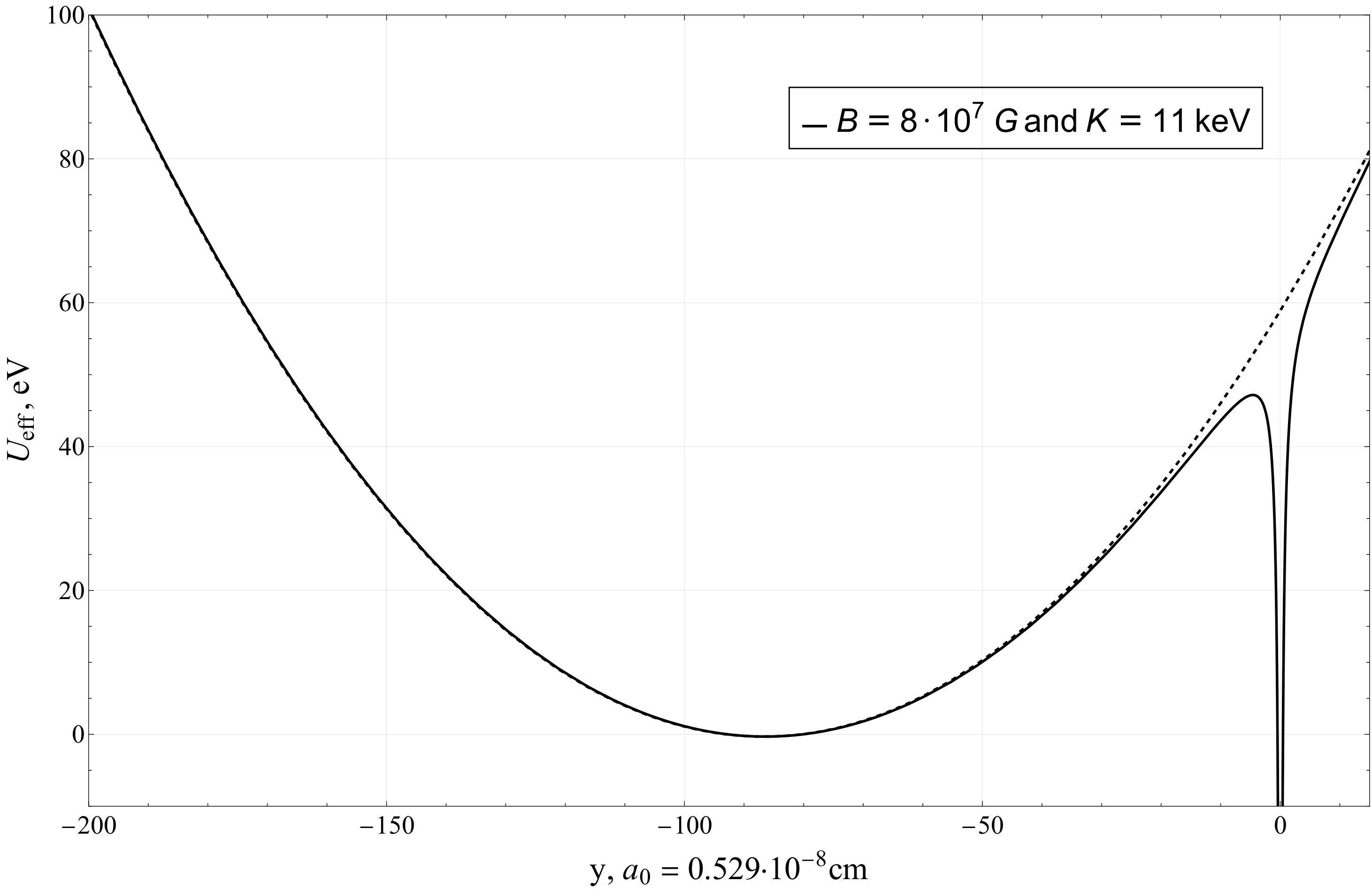}}
\caption{The illustrative plot of the oscillator approximation. The solid line is the projection of $\text{U}_{\text{eff}}$ on the $x=z=0$ plane. The dashed line is the parabolic approximation.}\label{fig:2}
\end{figure}

Next we return to eqs.(\ref{eq:forz})--(\ref{eq:potforz1}). Inequalities (\ref{eq:inequal}) permit to take the square root out of the integral (\ref{eq:potforz1}) at $\boldsymbol{\rho} = 0$, i.e., at $y = 0$ in the projection under consideration. Then (\ref{eq:longforeb}) takes the form
\begin{equation}\label{eq:forz1}
	\frac{d^2f}{dz^2} + m \bigg( \text{E} - \frac{e \text{B}}{m} + \frac{e^2}{\sqrt{y_0^2 + z^2}} \bigg) f(z) = 0.
\end{equation}
The characteristic distance for the dynamics of $f(z)$ is $z \sim \ell$ while according to (\ref{eq:inequal}) $|y_0| \gg \ell$. Therefore the potential energy in (\ref{eq:forz1}) can be expanded in a series in powers of $z^2$. As a result (\ref{eq:forz1}) reduces to the linear oscillator equation, see Fig.\ref{fig:2}
\begin{equation}\label{eq:osceq}
	\frac{d^2f}{dz^2} + m \bigg( \text{E} - \frac{e \text{B}}{m} + \frac{e^2}{|y_0|} - \frac{e^2}{|y_0|^3}\frac{z^2}{2} \bigg) f(z) = 0.
\end{equation}

The energy levels are given by
\begin{equation}\label{eq:spec}
	\text{E}_n = \frac{e \text{B}}{m} - \frac{e^2}{|y_0|} + \bigg(n + \frac{1}{2} \bigg) \sqrt{\frac{2 e^2}{m |y_0|^3}}.
\end{equation}

The states in MW are weakly bound. As an illustration we take $\text{B} = \text{B}_{12} = 10^{12}$ G and evaluate the ground state energy using (\ref{eq:spec}). The field strength $\text{B}_{12}$ corresponds to $\text{K}_{c} = 67$ keV, $\text{I} = 11.6$ keV, $\ell = 13 \text{ MeV}^{-1} \simeq a_0/20$. According to (\ref{eq:inequal}) the separated MW is formed when $|y_0| = \text{K}/e\text{B} \gg a_0$. In line with this condition we take $|y_0| = 10 a_0$, i.e., $\text{K} = 15$ MeV. Then (\ref{eq:spec}) yields 
\begin{equation}\label{eq:eexample}
    \text{E}_{\text{B}} = \text{E}_{n = 0} - \text{I} = -2.1 eV.
\end{equation}
This value is small compared with Ps ground state binding energy without MF equal to $-6.8$ eV and much smaller than the ground state binding energy of Ps at rest under the same $\text{B}_{12}$ MF which will be calculated in Sec.\ref{sec:6}.


Two remarks are relevant at this stage. Imposing the condition $(\text{E}_n - e\text{B}/m)<0$ we conclude that (\ref{eq:spec}) is valid up to
\begin{equation}
	n \lesssim \sqrt{\frac{m e^2 \text{K}}{2 (e \text{B})}} = \sqrt{\frac{|y_0|}{a_{\text{B}}(\text{Ps})}}.
\end{equation}
Beyond this restriction there is a condensation of infinite number of levels close to the ionization threshold. The transition from bound to autoionization states in hydrogen has been studied in \cite{19, 33, 34}. How to reframe the results of these works for Ps is an open question.

\section{The Overlapping Potential Wells}\label{sec:6}

As already explained, the configuration of the potential depends upon the values of the four basic parameters: $\text{K}, a_{\text{B}}(\text{Ps}), \ell, |\boldsymbol{\rho}_0| = |y_0|$ and on their dimensionless ratios. Under the condition (\ref{eq:inequal}) the outer well is formed, the adiabatic approximation is applicable and the two wells are isolated. Now we wish to consider the situation when the outer well is formed, the adiabatic approximation is valid but the two wells overlap. This corresponds to the following conditions
\begin{equation}\label{eq:inq}
	\text{K} > \text{K}_c, ~~~~~ \ell \ll a_{\text{B}}(\text{Ps}), ~~~~~ |y_0| \ll a_{\text{B}}(\text{Ps}).
\end{equation}
Note that we are not comparing $|y_0|$ and $\ell$ since their ratio depends on the value of $\text{K}$. In this section we partly rely on methods developed in \cite{13, 35, 36, 37, 38, 39, 40}.

Our task is to solve eq.(\ref{eq:forz}) under the conditions (\ref{eq:inq}). Firstly, we note that in the adiabatic approximation the size of Ps in the transverse plane is determined by the wave function (\ref{eq:R}) and is of the order of $\rho \sim \ell \ll a_{\text{B}}(\text{Ps})$. Therefore the density distribution of $|\varphi_{\mathbf{K}}(\boldsymbol{\rho}, z)|^2$ takes the elongated shape in $z$-direction. This gives birth to the idea to replace the potential (\ref{eq:potforz1}) by a MF independent one-dimensional potential. At $|z| \rightarrow \infty$ (\ref{eq:potforz1}) has the following asymptotic form
\begin{equation}
	\text{U}(z) \rightarrow \frac{e^2}{|z|} + \mathcal{O}\bigg( \frac{|y_0|^2}{|z|^3}\bigg).
\end{equation}

For (\ref{eq:forz}) with the account of (\ref{eq:defeb}) this yields
\begin{equation}\label{eq:onedim}
	\bigg( - \frac{1}{m} \frac{d^2}{dz^2} - \frac{e^2}{|z|} \bigg) f(z) = \text{E}_{\text{B}} f(z).
\end{equation}
This is a one-dimensional Schrödinger equation with a Coulomb potential studied in \cite{35, 36, 37, 38, 39, 40}. The difficulty to solve (\ref{eq:onedim}) arises from the pole in the potential at $z = 0$. The pole $1/|z|$ is more serious than $1/r$ in three dimensions because in the last case the singularity is largely neutralized by the space volume proportional to $r^2$. As a pure mathematical problem, eq.(\ref{eq:onedim}) does not have a complete set of solutions \cite{36}. There is a complete set of well-behaved odd-parity solutions, but the well-behaved even-parity solutions are absent \cite{35, 36, 38}. It is not evident how to match the even-parity solutions at the origin for the regions $z > 0$ and $z < 0$. One can speak of a «non-penetrability» of the one-dimensional Coulomb potential at $z = 0$. In terms of the general theory of differential equations the operator $[(-1/m)d^2/dz^2 - e^2/|z|]$ is not a self-adjoint one \cite{38}. Here we are interested in the ground state solution which is obviously an even-parity. A way to overcome the problem at the origin is to insert a cut-off at small distances \cite{35}. We follow an alternative approach of \cite{37, 39}. Eq. (\ref{eq:onedim}) is solved both inwards from $|z|=\infty$ and outwards from $z = 0$ matching the logarithmic derivatives of the two solutions at $|z| = z_0 \gtrsim \ell$. This procedure leads to an equation for the energy spectrum.


It is convenient to change the variable $z$ to $\xi = 2z/\nu a_{\text{B}}$ \cite{35} with $\nu$ and $\xi$ being dimensionless. Whereupon (\ref{eq:onedim}) reduces to
\begin{equation}\label{sec6:eq40}
    \frac{d^2}{d \xi^2} f(\xi) + \bigg[ \frac{1}{4} \bigg(m \nu^2 a^2_{\text{B}}\bigg)\text{E}_{\text{B}} + \frac{\nu}{|\xi|} \bigg]f(\xi) = 0,
\end{equation}
or 
\begin{equation}\label{eq:whitteq}
	\frac{d^2}{d\xi^2}f(\xi) + \bigg(-\frac{1}{4} + \frac{\nu}{|\xi|}\bigg) f(\xi) = 0,
\end{equation}
with $\nu$ related to the binding energy according to
\begin{equation}\label{eq:eb}
	\text{E}_{\text{B}} = - \frac{1}{m a_{\text{B}}^2 \nu^2}.
\end{equation}
Note that $\nu = 1$ corresponds to $\text{E}_{\text{B}} = \text{E}_0 \simeq - 6.8$ eV the ground state energy of Ps. Eq. (\ref{eq:whitteq}) is the Whittaker's confluent hypergeometric equation \cite{41}. The solution with positive $\nu$ that decreases exponentially at infinity is $f(\xi)= \text{const} \cdot W_{\nu, \sfrac{1}{2}}(\xi)$. To match with the interior solution it suffices to keep the first order in $z$ terms. For $0 < \nu \ll 1$ one can write \cite{37}
\begin{equation}\label{eq:Whittakersol}
	W_{\nu, \sfrac{1}{2}}\bigg(\frac{2 z}{\nu a_{\text{B}}}\bigg) = -1 + \frac{2 z }{a_{\text{B}}} \bigg( \ln \frac{2 z}{a_{\text{B}}} + \frac{1}{2 \nu} + \mathcal{C} - 1\bigg), 
\end{equation}
where $\mathcal{C}$ is the Euler constant ($\mathcal{C} = 0.5772...$). The logarithmic derivative reads
\begin{eqnarray}\label{eq:etaext}
	\eta(\nu, z)_{\text{ext}} &=& \frac{\partial}{\partial z} \ln W_{\nu, \sfrac{1}{2}}\bigg( \frac{2 z}{\nu a_{\text{B}}}\bigg) = \nonumber \\ & & = - \frac{1}{\nu a_{\text{B}}} - \frac{2}{a_{\text{B}}}\bigg( \ln \frac{2 z}{\nu a_{\text{B}}} + \mathcal{C} \bigg).
\end{eqnarray}

Next we consider the inner solution. The potential given by (\ref{eq:potforz1}) is an even function of $z$ and hence the solutions may be either even or odd. We focus on even solutions leaving the odd ones for another publication. Following \cite{37} we present the potential (\ref{eq:potforz1}) in the following form
\begin{eqnarray}
	\text{U}(z) = \frac{e^2}{2 \pi \ell^2}\iint d\boldsymbol{\rho}\frac{\exp(-\boldsymbol{\rho}^2/2\ell^2)}{\sqrt{(\boldsymbol{\rho}+\boldsymbol{\rho}_0)^2 + z^2}} = \nonumber \\\label{eq:upotentialstar} = \frac{e^2}{\ell}\vartheta\bigg(\frac{|z|}{\ell}\bigg),
\end{eqnarray}
where $\vartheta(\xi)$ is analytic for $|\xi| < \infty$. Introducing the variable $t = z/\ell$ we rewrite eq. (\ref{eq:longforeb}) in the following form
\begin{equation}\label{eq:fort}
	\frac{d^2}{dt^2}f(t) + \bigg( \frac{2 \ell}{a_{\text{B}}}\vartheta(t) - \frac{\ell^2}{a_{\text{B}}^2 \nu^2} \bigg) f(t) = 0,
\end{equation}
where the expression (\ref{eq:eb}) for $\text{E}_{\text{B}}$ has been used. Unlike the two and three dimensions for a one-dimensional Coulomb potential $\text{U}(|z|) = - e^2/|z|$ there is no continuity of the wave function and its derivative at $z = 0$. Therefore in zero order we take $\ell = 0$ (the infinite MF). In this case there are two solutions of (\ref{eq:fort}) 
\begin{equation}
	f^{(0)}(t) = \text{const}, ~~~~~ f^{(0)}(t) = \text{const} \cdot t.
\end{equation}
The first one corresponds to even solution. To first order in $\ell$ we have the equation 
\begin{equation}
	\frac{d^2}{dt^2}f(t) + \frac{2 \ell}{a_{\text{B}}} \vartheta(t)f^{(0)}(t) = 0.
\end{equation}
The solution reads
\begin{equation}
	f(t) = \text{const} \bigg( 1 - \frac{2 \ell}{a_{\text{B}}} \int\limits_{0}^{t} dt_1 \int\limits_{0}^{t_1}dt_2 \vartheta(t_2)\bigg).
\end{equation}
For $f^{(0)}(t) = \text{const}$, i.e., for even states, the solution is finite and non-vanishing at $z = 0$. Its logarithmic derivative is
\begin{equation}\label{eq:innersol}
	\eta(z)_{\text{int}} = - \frac{2}{a_{\text{B}}} \int\limits_{0}^{z/\ell}\vartheta(t)dt,
\end{equation}
where according to (\ref{eq:upotentialstar})
\begin{equation}\label{eq:vartheta}
	\vartheta(t) = \frac{1}{2 \pi \ell} \iint d\boldsymbol{\rho} \frac{\exp(-\boldsymbol{\rho}^2/2\ell^2)}{\sqrt{(\boldsymbol{\rho} + \boldsymbol{\rho}_0)^2 + \ell^2t^2}}.
\end{equation}
This is the point where the dependence on pseudomomentum $\textbf{K}$ comes into play through $\boldsymbol{\rho}_0 = - [\textbf{B}\times \textbf{K}]/e \text{B}^2$, or $y_0 = - \text{K}/e\text{B}$, $\textbf{K} = (\text{K}, 0, 0)$, see (\ref{eq:sep_pot_wells}) above. Insertion of (\ref{eq:vartheta}) into (\ref{eq:innersol}) yields 
\begin{equation}
	\eta(z)_{\text{int}} = - \frac{2}{a_{\text{B}}} \iint d \boldsymbol{\rho} \frac{\exp(-\boldsymbol{\rho}^2/2\ell^2)}{2 \pi \ell^2} \int\limits_{0}^{z/\ell} \frac{dt}{\sqrt{u^2 + t^2}},
\end{equation}
where $u^2 = (\boldsymbol{\rho}+\boldsymbol{\rho}_0)^2/\ell^2$. Integrating over $dt$ on obtains
\begin{eqnarray}
    \eta(z)_{\text{int}} = - \frac{2}{a_{\text{B}}} \iint d \boldsymbol{\rho} \frac{\exp(-\boldsymbol{\rho}^2/2\ell^2)}{2 \pi \ell^2} \bigg[ \ln\bigg(1 + \nonumber \\ \label{def:int} + \sqrt{1 + \frac{(\boldsymbol{\rho}+\boldsymbol{\rho}_0)^2}{z^2}}\bigg) + \ln \frac{z}{|\boldsymbol{\rho}+\boldsymbol{\rho}_0|}\bigg] = \mathcal{J}_1 + \mathcal{J}_2,
\end{eqnarray}
with $\mathcal{J}_1$ and $\mathcal{J}_2$ corresponding to the contribution of the first and second terms inside the square brackets. In the strong field limit $\ell \rightarrow 0$ and for $z/\ell \rightarrow \infty$ the integration of the first term is straightforward and gives $\mathcal{J}_1 \simeq -(2/a_{\text{B}})\ln2$. Evaluation of $\mathcal{J}_2$ is more complicated and is presented in the Appendix \ref{app:a}. Summing the two contributions we obtain
\begin{eqnarray}\label{eq:etaint}
    \eta(z)_{\text{int}} = - \frac{2}{a_{\text{B}}}\bigg( \ln \frac{z}{\ell} - \frac{1}{2} \ln 2 + \frac{1}{2} \mathcal{C} - \nonumber \\ - \frac{1}{2} \int\limits_{0}^{x} dt \exp(-t)\ln \frac{x}{t}\bigg|_{x = \frac{\ell^2 \text{K}^2}{2}} \bigg).
\end{eqnarray}

Equating the logarithmic derivatives (\ref{eq:etaext}) and (\ref{eq:etaint}) we arrive at the equation for $\nu$ determining according to (\ref{eq:eb}) the Ps ground state binding energy
\begin{equation}\label{eq:energyspec}
    \frac{1}{\nu} + \ln \frac{1}{\nu^2} = \ln \frac{a_0^2}{\ell^2} + \ln 2 - \mathcal{C} - \Lambda\bigg(\frac{\ell^2 \text{K}^2}{2}\bigg),
\end{equation}
where 
\begin{equation}\label{eq:energyspec1}
    \Lambda(x) = \int\limits_{0}^x dt \exp(-t) \ln \frac{x}{t}.
\end{equation}
The dependence on pseudomomentum in the form of the integral $\Lambda$ in (\ref{eq:energyspec}), (\ref{eq:energyspec1}) was given in \cite{13} with a reference to \cite{37} though in \cite{37} the reader does not find an explicit dependence on pseudomomentum. The actual calculations of $\eta(z)_{\text{int}}$ which include the dependence on the pseudomomentum are given by (\ref{eq:etaint}) and (\ref{eq:a1})--(\ref{eq:a4}). Expansion of $\Lambda$ in different limiting cases is presented in the Appendix \ref{app:b}. At $x \rightarrow 0$ the function $\Lambda(x)$ behaves as $\Lambda(x)\simeq x + \mathcal{O}(x^2)$. At $x\rightarrow\infty$ one has $\Lambda(x) \simeq \ln x + \mathcal{C} + \mathcal{O}(e^{-x})$. The derivation of (\ref{eq:energyspec}) is based on the expansion (\ref{eq:Whittakersol}) valid for $\nu \ll 1$ which corresponds to the ground state. The excited even and odd states will be the subject of the forthcoming publication.

\begin{figure}
\center{\includegraphics[width = 1.0 \linewidth]{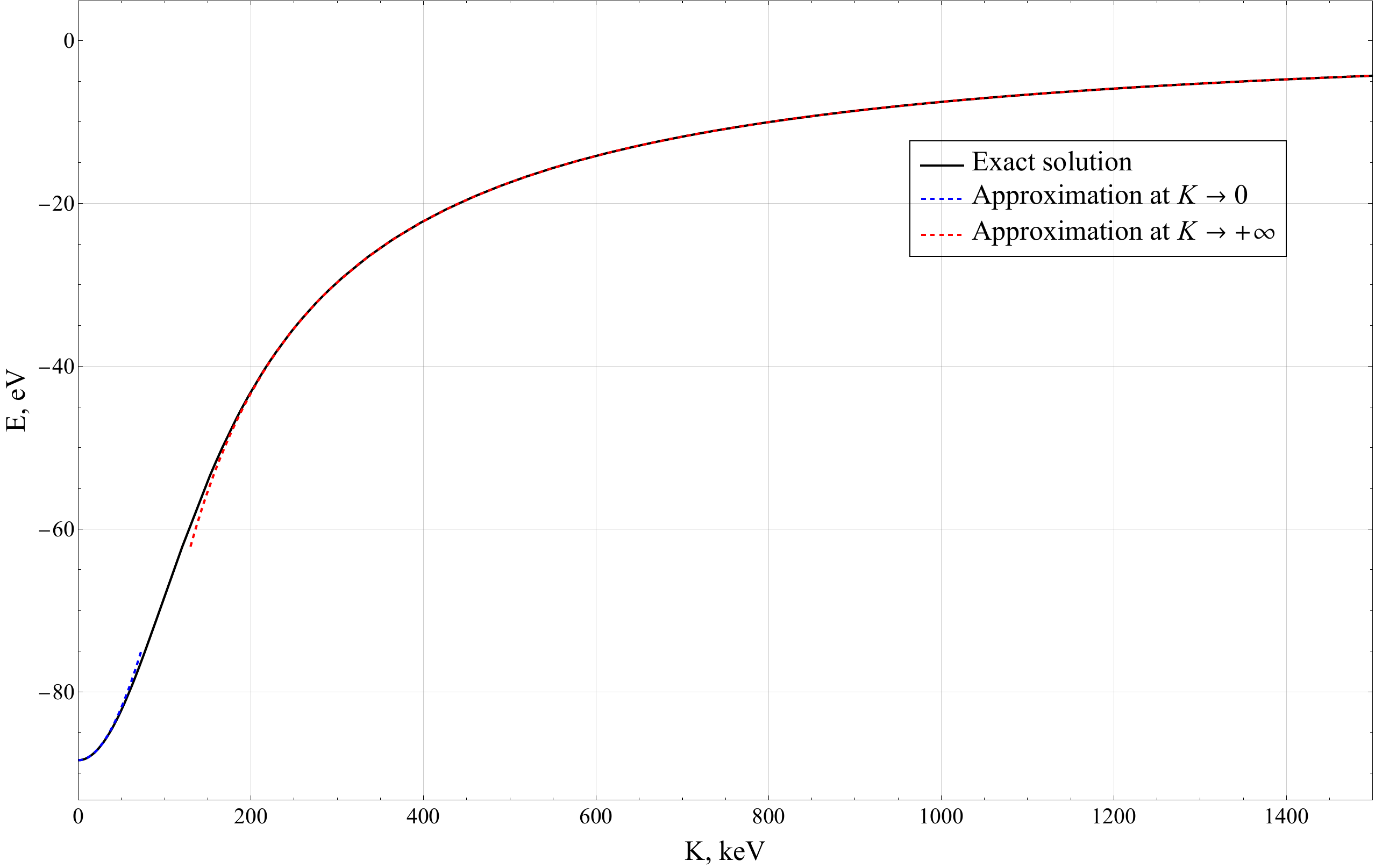}}
\caption{The dependence of the binding energy for the ground state (\ref{eq:eb}) on the pseudo-momentum K. The blue dashed line corresponds to the limiting case of small K (\ref{b:limit_small_x}), the red line corresponds to the limiting case of large K (\ref{b:limit_big_x}) (see Appendix \ref{app:b}). $\text{B} = \text{B}_{12} = 10^{12}$ G.}\label{fig:3}
\end{figure}

The dependence of the ground state binding energy on pseudomomentum for $\text{B} = \text{B}_{12}$ is shown in Fig.~\ref{fig:3}. Equation (\ref{eq:energyspec}) corresponds to the set of parameters meeting the conditions (\ref{eq:inq}). Therefore $\text{K}$ can not be arbitrary large, namely $\text{K}\ll2a_0(e\text{B}_{12})\simeq3$ MeV. The condition $x = \text{K}^2\ell^2/2\ll1$ is more stringent, $\text{K}\ll0.1$ MeV. Nevertheless, in Fig.~\ref{fig:3} we show the $\text{E}_{\text{B}}(\text{K})$ dependence up to $\approx 1.5$ MeV to show the continuous transition from the overlapping wells regime to the outer well configuration. However, the continuous transformation from the distorted Coulomb solution through the saddle states into the MW can be followed only by the numerical calculations \cite{28}. For $\ell^2\text{K}^2\ll1$ the term $\Lambda$ in (\ref{eq:energyspec}) gives the following contribution to the ground state binding energy 
\begin{equation}
    \text{E}_{\text{B}} - \text{E}_{\text{B}0} = |\text{E}_{\text{B}0} - \text{E}_{\text{B}}| = \frac{\ell^2 \text{K}^2}{4 m a_0^2 \nu_0},
\end{equation}
where $\text{E}_{\text{B}0}$ and $\nu_0$ corresponds to (\ref{eq:energyspec}) with $\Lambda \equiv 0$.

Before we give further consideration to the equation (\ref{eq:energyspec}), we have to look at the issue of the gauge choice in the Ps problem. The reason is that the attention to the role of Ps in the NS magnetosphere was drawn by the pioneering works \cite{8, 9} performed in the Landau gauge.

\section{The Issue of the Gauge Choice in the Ps Problem}\label{sec:7}

It is well known that the problem of a charged particle or of a neutral two-particle system moving in a constant MF may be solved either in a symmetric gauge or in the Landau one \cite{30}. The symmetric gauge which we adopt in the present work has been commonly used in treating the neutral two-body systems in MF, see the references cited above and \cite{1046} as a possible starting point. On the other hand, the problem of Ps moving in the magnetosphere of a NS has been studied using the Bethe-Salpeter equation in the Landau gauge \cite{8, 9, 10}. We would like to recapitulate the main points concerning the relation between the two gauges with the emphasis on the transformation properties of the pseudomomentum. The gauges under condition are 
\begin{eqnarray}
    \text{Symmetric (SG): } \mathbf{A}^{S} &=& \frac{1}{2}\text{B}(-y, x, 0), \\
    \text{Landau (LG): }\mathbf{A}^{L}&=& \text{B}(-y, 0, 0).
\end{eqnarray}

Sometimes $\mathbf{A}^{L}$ is called the 1st LG and the second LG is defined as $\mathbf{A}^{L2} = \text{B}(0, x, 0)$. The SG is related to the LG by the following gauge transformation 
\begin{eqnarray}\label{eq:gauge1}
    \mathbf{A}^{L} &=& \mathbf{A}^{S} + \boldsymbol{\nabla}f(x, y, z), \\ \label{eq:gauge2}
    \text{B} \begin{bmatrix} -y\\0\\0\end{bmatrix} &=& \frac{\text{B}}{2}\begin{bmatrix}
        -y\\x\\0
    \end{bmatrix} + \boldsymbol{\nabla} f(x, y, z).
\end{eqnarray}
From (\ref{eq:gauge1}) and (\ref{eq:gauge2}) one gets $f(x, y, z) = (-1/2)\text{B}xy$. The Ps Hamiltonian in an arbitrary gauge is given by (\ref{eq:Hamiltoniantwoparticles}). The pseudomomentum which commutes with $\hat{\mathcal{H}}$ and can be diagonalized simultaneously with $\hat{\mathcal{H}}$ reads \cite{13} 
\begin{eqnarray}\label{eq:Kanygauge}
    \hat{\mathbf{K}} = \hat{\mathbf{p}}_1 + \hat{\mathbf{p}}_2 &-& e (\mathbf{A}_1 - \mathbf{A}_2) + \nonumber\\ &+& e [\mathbf{B}\times (\mathbf{r}_1 - \mathbf{r}_2)].
\end{eqnarray}

The previously presented formula (\ref{eq:pseudomomentum}) corresponds to the SG. For the neutral two-body system the components $\hat{\mathbf{K}}_x$ and $\hat{\mathbf{K}}_y$ commute with one another. The expression (\ref{eq:Kanygauge}) is obviously gauge dependent and takes the following forms in SG and LG
\begin{eqnarray}
    \hat{\mathbf{K}}^S &=& \hat{\mathbf{p}}_1 + \hat{\mathbf{p}}_2 + \frac{1}{2}e\text{B}\begin{bmatrix}
        -(y_1 - y_2)\\x_1-x_2\\0
    \end{bmatrix},\\
    \hat{\mathbf{K}}^L &=& \hat{\mathbf{p}}_1 + \hat{\mathbf{p}}_2 + e\text{B}\begin{bmatrix}
        0\\x_1 - x_2\\0
    \end{bmatrix}.\label{eq:viewKL}
\end{eqnarray}
Worth noting that $\hat{\mathbf{K}}^L_x$ coincides with Ps canonical momentum $\mathbf{P}_x = \mathbf{p}_1 + \mathbf{p}_2$. One easily finds that $\hat{\mathbf{K}}^S$ and $\hat{\mathbf{K}}^L$ are related by the unitary gauge transformation
\begin{equation}\label{eq:Ktransformation}
    \hat{\mathbf{K}}^L = \mathcal{U} \hat{\mathbf{K}}^S \mathcal{U}^{\dagger},
\end{equation}
where 
\begin{equation}\label{eq:expu}
    \mathcal{U}(x_i, y_i) = \exp\bigg\{-\frac{ie\text{B}}{2} \bigg(x_1y_1 - x_2 y_2\bigg)\bigg\}, i = 1, 2.
\end{equation}

The next step will be to show that the expectation values of the pseudomomentum components over the wave function are gauge-invariant quantities. The gauge transformation of the pseudomomentum is compensated by the corresponding transformation of the wave function. This is an anticipated results since $\hat{\mathbf{K}}$ and $\hat{\mathcal{H}}$ can be diagonalized simultaneously. From now onwards we make the projection onto the $x = z = 0$ plane and $\hat{\mathbf{K}} = (\hat{\mathbf{K}}_x, 0, 0)$. The corresponding arguments have been presented prior to (\ref{eq:effpot1}) and following (\ref{eq:sep_pot_wells}). We introduce the wave functions $|\psi^S\rangle$ and $|\psi^L\rangle$ in SG and LG correspondingly \cite{30}. The function $|\psi^S\rangle$ is the simultaneously eigen-function of the Hamiltonian and the operator $\hat{\mathbf{K}}_x$
\begin{eqnarray}\label{eq:spectraleq1}
    \hat{\mathcal{H}}|\psi^S\rangle &=& \text{E}|\psi^S\rangle,\\
    \hat{\mathbf{K}}_x^{S}|\psi^S\rangle &=& \text{K}_x^S |\psi^S\rangle,\label{eq:spectraleq2}
\end{eqnarray}
where $\text{K}_x$ is the eigenvalue of $\hat{\mathbf{K}}_x$, i.e., the $x$-component of Ps pseudomomentum. Multiplying both sides of (\ref{eq:spectraleq2}) by $\mathcal{U}$ from the left and inserting the identity $\mathcal{U}^{\dagger}\mathcal{U} = \mathbbm{1}$, one obtains
\begin{equation}
\bigg(\mathcal{U}\hat{\mathbf{K}}_x^S\mathcal{U}^{\dagger}\bigg)\mathcal{U}|\psi^S\rangle = \text{K}_x^S\mathcal{U}|\psi^S\rangle.
\end{equation}

Equation (\ref{eq:Ktransformation}) and the relation 
\begin{equation}\label{eq:simpeq}
    \mathcal{U}|\psi^S\rangle = |\psi^L\rangle
\end{equation} 
allow to recast this equation as follows
\begin{equation}\label{eq:eigenKS}
    \hat{\mathbf{K}}^L_x |\psi^L\rangle = \text{K}_x^S|\psi^L\rangle.
\end{equation}
On the other hand, $\hat{\mathbf{K}}_x$ is diagonalized in LG representation as 
\begin{equation}\label{eq:eigenKL}
\hat{\mathbf{K}}_x^L|\psi^L\rangle = \text{K}_x^L|\psi^L\rangle.
\end{equation}
Comparing (\ref{eq:eigenKS}) and (\ref{eq:eigenKL}) we conclude that the expectation value of $\hat{\mathbf{K}}_x$ is independent of the gauge choice $\text{K}_x^S = \text{K}_x^L = \text{K}_x$. We remind that according to (\ref{eq:viewKL}) $\text{K}_x^L = p_{1x} + p_{2x} = \text{P}_x$, where $\text{P}_x$ is the $x$-component of the Ps center-of-mass canonical momentum.

At this point a question may arise over the validity of the simple transformation relation (\ref{eq:simpeq}). The detailed analysis of the wave functions gauge transformation is missing in most textbooks. In \cite{30} it is explained that the gauge transformation of the wave function follows the transformation of the potential $\mathbf{A} \rightarrow \mathbf{A} + \boldsymbol{\nabla}f, |\psi\rangle \rightarrow \exp(i e f)|\psi\rangle$ which corresponds to (\ref{eq:gauge1}), (\ref{eq:spectraleq1}) and (\ref{eq:simpeq}). The problem of the wave function gauge transformation has a long history \cite{231, 135, 136, 137, 138} and is under discussion up to now \cite{139, 140}. The gauge connection is nontrivial due to the degeneracy of the wave functions in both gauges. It has been shown \cite{136, 137, 138, 139, 140} that a «proper» gauge transformation should include the simultaneous diagonalization of the pseudomomentum operator along with the Hamiltonian as given by (\ref{eq:spectraleq1})--(\ref{eq:spectraleq2}). This allows to choose the phases of $|\psi^S\rangle$ and $|\psi^L\rangle$ so that $|\psi^L\rangle = \mathcal{U}|\psi^S\rangle$. It is beyond the purpose of this work to dwell into this complicated subject. 

The conclusion is that the expectation value of the pseudomomentum $\text{K}_x$ is gauge independent and it reduces in LG to the canonical momentum used in \cite{10, 998}. It means that Ps binding energy should be the same for equal values of $\text{K}_x$ and $\text{P}_x$. If this is not the case, one should look into calculations.

\section{The Ground State Energy Equation: Accurate as Approximate}\label{sec:8}

As stated in the Introduction and in the previous Section, the current interest in Ps physics is to a great extent due to its possible role in processes taking place in NS magnetosphere \cite{8, 9, 10, 11}. The problem has been raised in the fundamental articles by A.E. Shabad and V.V. Usov \cite{8, 9}. In \cite{10} and \cite{45} the authors continued the investigation with the emphasis on the Ps contribution into the photon polarization operator \cite{998}. The ground state energy of Ps moving across MF was found using the Bethe-Salpeter equation in LG. In \cite{555} the same problem was solved using the Bethe-Salpeter equation in LG2, i.e., with $\mathbf{A}^{L2} = \text{B}(0, x, 0)$. Our task is to compare the results for the Ps ground state energy obtained in \cite{8, 9, 10, 45, 998} and in the present work. To be more concrete, we shall refer to the equations from \cite{10}. A comparison is inhibited by the different initial formulation of the problem. In the first case the research target is the polarization operator in strong MF. The Ps spectrum is given by the poles of this operator. In our study the starting point is the Hamiltonian (\ref{eq:Hamiltoniantwoparticles}) from which the one-dimensional equations (\ref{eq:forz}), (\ref{eq:longforeb}), (\ref{eq:onedim}) are derived in the adiabatic approximation. The important point is that the Bethe-Salpeter approach in the adiabatic approximation reduces to the one-dimensional Schrödinger equation with Coulomb potential ((3.21), (3.22) of \cite{10}). This makes it possible to compare the methods to solve this equation and the obtained results. After more than 60 years of investigations following the publication of Loudon \cite{35} with the cut-off solution of the singularity problem, the theory of one-dimensional hydrogen-like system in MF remains a topic of debate and controversy \cite{36, 765, 38}. In the present work the cut-off method is not used. Instead, the ground state eigenvalue equation (\ref{eq:energyspec}) has been derived by matching the logarithmic derivative of the outer and inner space wave functions. We remind that (\ref{eq:energyspec}) corresponds to the overlapping Coulomb and magnetic wells. The energy levels in the shallow outer well are described by (\ref{eq:spec}). In order to understand better the structure of the equation (\ref{eq:energyspec}) we consider first the case $\mathbf{K} = 0$. It might be tempting to take (\ref{eq:energyspec}) in the following truncated form 
\begin{equation}\label{sec8:eq73}
    \frac{1}{\nu} = \ln \frac{a_0^2}{\ell^2},
\end{equation}
where the $\ln 1/\nu^2$ contribution as well as $(\ln 2 - \mathcal{C}) = 0.116$ term have been omitted. According to (\ref{eq:eb}) the ground state energy is
\begin{equation}\label{eq:ebLL}
    \text{E}_{\text{B}} = - \frac{m e^4}{4}\ln^2\frac{a_0^2}{\ell^2}.
\end{equation}
This formula is a duplicate of the well-known equation from the Landau and Lifshitz \cite{30} for the hydrogen atom in MF (with the replacement of the reduced mass and Bohr radius). It is instructive to compare the results for $\text{E}_{\text{B}}$ given by (\ref{eq:ebLL}) with the numerical solution of (\ref{eq:energyspec}). For MF strength $\text{B}_{12} = 10^{12}$ G (\ref{eq:ebLL}) gives $\text{E}_{\text{B}} = -250$ eV while according to (\ref{eq:energyspec}) $\text{E}_{\text{B}} = - 88$ eV. For the hydrogen atom the poor accuracy of the $\ln^2-$type formulas is a firmly established fact \cite{38, 765}. The conclusion of \cite{38} based on the comparison with the numerical calculations is that such formulas have no range of applicability.

Next we turn to the $\text{K} > 0$ situation. The dependence of $\text{E}_{\text{B}}$ on the pseudomomentum $\text{K}$ according to (\ref{eq:energyspec}) is shown in Fig.~\ref{fig:3}. Using the notations of the present article, the ground state energy of Ps moving in MF obtained in \cite{10} (eq. 3.27) reads 
\begin{equation}\label{sec8:eq117}
    \text{E}_{\text{B}} = - \frac{m e^4}{4}\ln^2\frac{a_0^2}{\ell^2 (1 + \ell^2 \text{P}_x^2)}.
\end{equation}

According to \cite{10} the quantum numbers attributed to the ground state are $n = n' = n_c = 0$, where $n, n'$ are Landau numbers of $e^-$ and $e^+$, $n_c$ is the label of the Coulomb level. Temporarily ignoring the poor accuracy of the $\ln^2-$type formulas, we deduce an equation similar to (\ref{sec8:eq117}) from (\ref{eq:energyspec}). We take (\ref{eq:energyspec}) in the following truncated form
\begin{equation}\label{sec8:eq118}
    \frac{1}{\nu} \simeq \ln \frac{a_0^2}{\ell^2} - \Lambda\bigg(\frac{\ell^2 \text{K}^2}{2}\bigg).
\end{equation}
At $\ell^2 \text{K}^2/2\ll1$ one has $\Lambda(x)\simeq x \simeq\ln(1 + x)$. This leads to
\begin{equation}\label{sec8:eq119}
    \text{E}_{\text{B}} = - \frac{m e^4}{4} \ln^2\frac{a_0^2}{\ell^2(1 + \ell^2 \text{K}^2/2)}.
\end{equation}
The two equations (\ref{sec8:eq117}) and (\ref{sec8:eq119}) have similar structure but still differ from each other. In no way do we wish to state that (\ref{sec8:eq119}) is a correct one. In fact, both are highly inaccurate as pointed out above. It would be interesting to see the equation similar to (\ref{eq:energyspec}) which has lead to (\ref{sec8:eq117}) in \cite{10}. To this end we follow the derivation of (\ref{sec8:eq117}) in \cite{10}. In the adiabatic approximation the Bethe-Salpeter equation is eventually reduced to the Schrödinger equation with the potential ((3.21) of \cite{10})
\begin{equation}
    V_0(z_1-z_2) = - \frac{e^2}{\sqrt{(z_1 - z_2)^2 + \ell^4 \text{P}_x^2}}.
\end{equation}
To use this potential in the domain of small $(z_1 - z_2)^2 + \ell^4\text{P}_x^2$ a term $\ell^2$ is added which plays the role of a cut-off at $\text{P}_x\rightarrow0$. Finally, the potential is approximated by the expression ((3.26) of \cite{10})
\begin{equation}\label{sec8:eq121}
    V(z_1 - z_2) = - \frac{e^2}{|z_1 - z_2| + \sqrt{\ell^2 + \ell^4 \text{P}_x^2}}.
\end{equation}
with this potential the result (\ref{sec8:eq117}) for the ground state binding energy is written down with a reference to \cite{35} for the solution of the one-dimensional Schrödinger equation with a cut-off. The potential considered in \cite{35} is $V(x) = - e^2/(a + |x|)$, where $a$ is the cut-off parameter. In \cite{35} MF is not considered and the ground-state eigenvalue equation reads 
\begin{equation}\label{sec8:eq122}
    \frac{1}{\nu} + \ln \frac{1}{\nu^2} = \ln \frac{a_0^2}{4 a^2}.
\end{equation}

In \cite{35} the $\ln^2-$type solution like (\ref{eq:ebLL}), (\ref{sec8:eq117}), (\ref{sec8:eq119}) is not presented. According to (\ref{sec8:eq121}) the role of cut-off $a$ is played by $a = \sqrt{\ell^2 + \ell^4 \text{P}_x^2}$ so that (\ref{sec8:eq122}) leads to the eigenvalue equation similar to (\ref{eq:energyspec})
\begin{equation}
    \frac{1}{\nu} + \ln \frac{1}{\nu^2} = \ln \frac{a_0^2}{\ell^2 (1 + \ell^2 \text{P}_x^2)} - 2 \ln 2.
\end{equation}

Omitting $\ln1/\nu^2$ and $2 \ln 2$ terms one obtains the result (\ref{sec8:eq117}) of \cite{10}. Within the adopted accuracy the term $\ell^2$ added to $(z_1 - z_2)^2 + \ell^4 \text{P}_x^2$ can be replaced by $2 \ell^2$ leading to the equation 
\begin{equation}\label{sec8:eq124}
    \frac{1}{\nu} + \ln \frac{1}{\nu^2} = \ln \frac{a_0^2}{\ell^2 (1 + \ell^2 \text{P}_x^2/2)} - 3 \ln 2.
\end{equation}
This has to be compared with (\ref{eq:energyspec}) in the approximation $\ell^2 \text{K}^2/2\ll1$, $\Lambda(x)\simeq x\simeq \ln(1 + x)$
\begin{equation}\label{sec8:eq125}
    \frac{1}{\nu} + \ln \frac{1}{\nu^2} = \ln \frac{a_0^2}{\ell^2 (1 + \ell^2 \text{K}^2/2)} + \ln 2 - \mathcal{C}.
\end{equation}

The fact that (\ref{sec8:eq124}) and (\ref{sec8:eq125}) differ by a shift from each other is no surprise. In \cite{38} one finds a list of 5 analytical adiabatic approximations available in the literature for the hydrogen atom energy levels as a function of MF. The Loudon formula (\ref{sec8:eq122}) is considered as moderately accurate and the equation close to (\ref{eq:energyspec}) is regarding as approaching an exact solution.

As stated in the Introduction, the detailed numerical calculations of Ps properties in MF were performed in \cite{27, 28} in the MF range ($10^{-3}-100$)$\text{B}_a$ and the pseudomomentum interval $\text{K} = (0-70)$ keV. Our calculations are performed in the adiabatic approximations $\text{B} \gg \text{B}_a$. The agreement of our results with that of \cite{28} is fairly well. For $\text{B} = 100 \text{B}_a$, $\text{K} = 37$ keV TABLE \RomanNumeralCaps{2} of \cite{28} gives $\varepsilon_g = 2.652$ keV which in our definition (\ref{eq:defeb}) $\text{E}_{\text{B}} = \varepsilon_g - \text{I}$ corresponds to $\text{E}_{\text{B}} = - 69$ eV. Our equation (\ref{eq:energyspec}) gives $\text{E}_{\text{B}} = -42$ eV while (\ref{sec8:eq117}), i.e. (3.27) of \cite{10} yields $\text{E}_{\text{B}} = -104$ eV.



\section{Implications}\label{sec:9}

We have shown that the Ps wave function and the energy spectrum experience deep transformation as it moves through a MF. This carries far-reaching physical implications. 

\begin{figure}
\center{\includegraphics[width = 1.0 \linewidth, keepaspectratio]{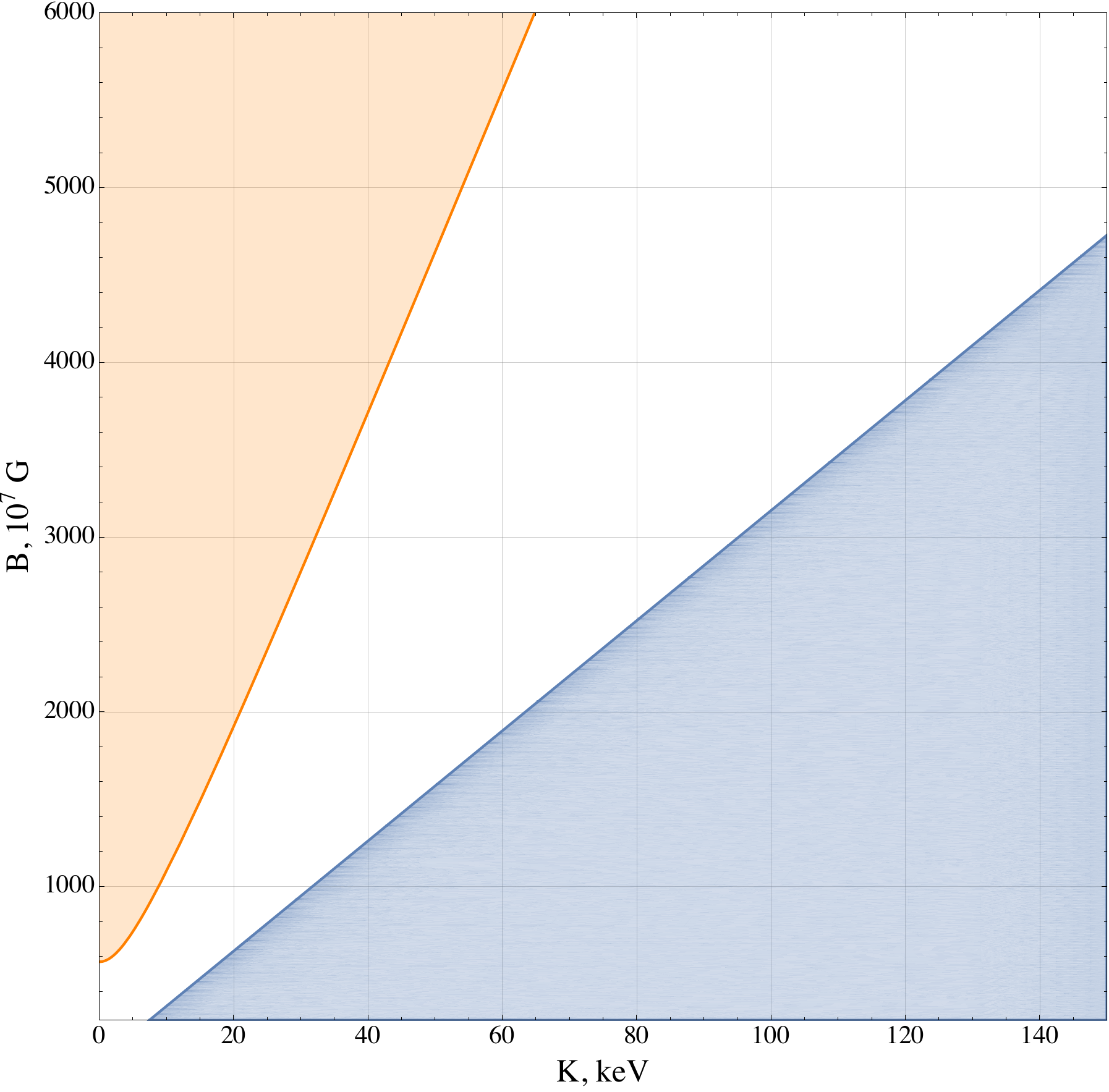}}
\caption{Phase space of parameters ($\text{K}, \text{B}$) for a system in a magnetic field. The orange region corresponds to the «merging» of two potential wells, the blue region corresponds to separated wells.}\label{fig:4}
\end{figure}

The first problem to consider is the one-photon Ps annihilation. It is a matter of a long standing dispute started by the work of Carr and Sutherland \cite{44}. The authors concluded that this process is possible from the Ps ground state and calculated its rate. This point was disproved by Wunner and Herold \cite{24}. They advocated the assumption that the energy minimum corresponds to $\mathbf{K} = 0$ state and proved that the one-photon annihilation from this state is impossible. The same conclusion is shared in \cite{10, 998} where it is stated that the one-photon annihilation can proceed from $\text{P}_x > 0$ states, where $\text{P}_x$ is the pseudomomentum in LG. At this point one can raise a question of whether the $\mathbf{K} = 0$ state from which the one-photon annihilation is forbidden is always a global ground state of Ps. We have seen that under the conditions (\ref{eq:inequal}) the outer MW emerges with the spectrum of the energy levels given by (\ref{eq:spec}). On the other hand, with $\text{K}$ increasing above $\text{K}_c$ the states in the CW are pushed out of the well into the continuum \cite{28}. In that way the ground state of the MW with $\text{K}\gg\text{K}_c$ becomes the global ground state of Ps. This is illustrated by the phase diagram on the «magnetic field -- pseudomomentum» plane shown in Fig. \ref{fig:4}. The orange region corresponds to the solution of eq. (\ref{eq:forz}), (\ref{eq:onedim}) under the conditions (\ref{eq:inq}), i.e., to the overlapping wells. Within the blue region the system is described by eq. (\ref{eq:forz1}) under the conditions (\ref{eq:inequal}), and this is the regime of separated potential wells. The two regions do not intersect which indicates the presence of the transition, or saddle, region which can not be investigated analytically. The migration of the ground state from the CW via saddle states towards the MW corresponds to the motion along the horizontal line in the direction of the increasing $\text{K}$. At $\text{K} \gg \text{K}_c$ the wave function relocates to the wide and shallow MW. Accordingly, the square of the wave function $|\psi(0)|^2$ evaluated at contact becomes near zero giving rise to the long-lived Ps ground state \cite{27, 28}. To put this statement on a quantitative basis we note that for the MF strength $\text{B}\gg\Delta/4\mu_0 = 3.63$ T Ps has a maximum symmetry spin state with spin-down $e^-$ and spin-up $e^+$ \cite{7, 30}. Here $\Delta = 8.4 \cdot 10^{-4}$ eV is the hyperfine splitting, $\mu_0 = 5.79 \cdot 10^{-5}$ eV/T is the Bohr magneton. Therefore it s sufficient to consider only the coordinate part of the Ps wave function. According to (\ref{eq:phi})--(\ref{eq:R}) the square of the ground state wave function at $\boldsymbol{\eta} = \mathbf{r}_1 - \mathbf{r}_2 = 0$, i.e., at $\boldsymbol{\rho}' = - \boldsymbol{\rho}_0$, $z = 0$ reads 
\begin{eqnarray}
    |\varphi_{\mathbf{K}}(0)|^2 & = & \frac{1}{2 \pi \ell^2} \exp\bigg(-\frac{\boldsymbol{\rho}_0^2}{2 \ell^2}\bigg) f^2(0) = \nonumber\\ & & = \frac{1}{2 \pi \ell^2}\exp\bigg(-\frac{\text{K}^2 \ell^2}{2}\bigg) f^2(0).
\end{eqnarray}
For the separated potential wells $f(z)$ satisfies the oscillator equation (\ref{eq:osceq}). The ground state solution is 
\begin{equation}
    f(z) = \bigg(\frac{m \omega}{\pi}\bigg)^{\sfrac{1}{4}}\exp\bigg(-\frac{m \omega}{2}z\bigg),
\end{equation}
where
\begin{equation}
    \omega^2 = \frac{2 e^2}{m|y_0|^3} = \frac{2 e^2}{m \text{K}^3 \ell^6}.
\end{equation}
Therefore 
\begin{equation}\label{eq:phizero}
    |\varphi_{\textbf{K}}(0)|^2 = \frac{1}{2 \pi \ell^2} \sqrt{\frac{m \omega}{\pi}} \exp\bigg(- \frac{\text{K}^2 \ell^2}{2}\bigg).
\end{equation}
This has to be compared with $|\varphi(0)|^2 = 1/\pi a_{\text{B}}^3(\text{Ps})$ for Ps at rest without MF. From (\ref{eq:phizero}) it follows that the probability density at the origin exponentially drops for $\text{K}^2\ell^2/2\gg1$. For $\text{B}_{12} = 10^{12}$ G and $\text{K} = 15$ MeV which fits the conditions (\ref{eq:inequal}) one has $\ell^2\text{K}^2/2 = 2 \cdot 10^4$ so that the $\exp\{-\ell^2\text{K}^2/2\}$ prevents Ps from annihilation for practically infinite time.

Another important property of the Ps decentered configuration is that it has a giant electric dipole moment \cite{13, 27, 28, 42, 43}. From the physical considerations it is clear that the average distance between particles if given by the difference (\ref{eq:gyromotion}) between the coordinates of the gyro-motion guiding center of $e^+$ and $e^-$
\begin{equation}
    \langle \boldsymbol{\eta}\rangle = \frac{\mathbf{K}_1 \times \mathbf{B}}{e \mathbf{B}^2} - \frac{\mathbf{K}_2 \times \mathbf{B}}{-e \mathbf{B}^2} = - \frac{\mathbf{B}\times \mathbf{K}}{e \mathbf{B}^2}.
\end{equation}
Repeating the arguments that led to (\ref{eq:effpot1}), consider the gauge $\text{K}_y = 0, \mathbf{K} = (\text{K}, 0, 0)$. Then $\langle \boldsymbol{\eta}\rangle_y = y_0 = - \text{K}/e \text{B}$ and 
\begin{equation}
    d = e y_0 = - \frac{\text{K}}{\text{B}}.
\end{equation}

As noted before, and as it is clear from Fig.\ref{fig:2}, $y_0$ coincides with the position of the MW minimum at $\text{K}\gg\text{K}_c$. According to (\ref{eq:inequal}) in the configuration with the decentered MW $|y_0| \gg a_{\text{B}}(\text{Ps})$. It means that the dipole momentum can exceed by many orders of magnitude the value corresponding to the atomic unit of length. The Ps with a giant dipole moment plays an important role in the NS magnetosphere at the polar gap \cite{42}. The electric field exerts a torque on the Ps dipole and causes it to rotate with angular velocity proportional to d. According to \cite{42} this rotation prevents the increase of the distance between $e^+$ and $e^-$ and thereby prevents the ionization of Ps inside the polar gap.

\section{Summary and Future Prospects}\label{sec:10}

We have investigated the transformation of the Ps wave function and energy spectrum taking place when Ps is moving through a MF. The main difference from the unmoving case is that the internal dynamics can not be separated from the centre-of-mass motion. The spectrum and the wave function parametrically depend on the generalized momentum operator $\mathbf{K}$ also called pseudomomentum. When $\text{K}$ exceeds the critical value (\ref{critpseudomomentum}) the outer «magnetic» potential well is formed in addition to the distorted Coulomb one. Two principal configurations are possible:
        \begin{enumerate}
            \item[(i)] The separated Coulomb and outer potential wells with a potential barrier inbetween.
            \item[(ii)] The overlapping wells with bilocalized wave function.
        \end{enumerate}
Both cases correspond to $\text{K} > \text{K}_c$ and to MF strong enough for adiabatic approximation to be legitimate
\begin{equation}
    \gamma = \frac{\text{B}}{\text{B}_a} = \frac{\omega_c}{2 R_y} \gg 1.
\end{equation}
Strong MF with $\gamma \gg 1$ is of a high astrophysical relevance. Which of the two above options is realized depends on the relations (\ref{eq:inequal}) and (\ref{eq:inq}) between the three basic parameters: the Bohr radius $a_{\text{B}}(\text{Ps}) = 2/me^2$, the Landau radius $\ell = (e\text{B})^{-1/2}$, and the difference $\text{K}/e\text{B}$ between the coordinates of the gyro-motion guiding centers of $e^+$ and $e^-$. With $\text{K}$ continuously increasing above $\text{K}_c$ the former Coulomb states are pushed above the ionization threshold and the spectrum resides in the decentered outer well. In this case the ground state of the MW becomes the energetically global ground state of Ps. The states in the MW have infinitesimal annihilation rates and giant dipole moments.

We solved the problem in the symmetric gauge which is commonly used in treating the neutral two-body systems in MF. Previously the ground state energy of Ps moving across MF was found using the Bethe-Salpeter equation in the Landau gauge \cite{8, 10, 998, 555}. In order to compare the results we reverted back to the old problem of the transformation between the two gauges with the primary interest in the pseudomomentum transformation. Pseudomomentum is taken as an additional operator which must be diagonalized along with the Hamiltonian in both gauges. The natural conclusion is that the expectation value of the pseudomomentum is gauge independent. It means that the Ps binding energy should be the same for equal values of the pseudomomentum.

An interesting problem which was left aside in our work is magnetically stimulated Ps center-of-mass chaotic diffusion motion \cite{47, 48, 49}. Roughly speaking, the center-of-mass undergoes a transition from regular to Brownian motion if the internal motion changes from regular to chaotic. Worth mentioning that diffusion motion may be important for high precision experiments with antihydrogen atoms \cite{50}.




\section{Acknowledgments}

The authors would like to thank M.A. Andreichikov for useful remarks and discussion.

\appendix

\section{Calculation of the integral $\mathcal{J}_2$}
\label{app:a}
Let's transform the original expression for $\mathcal{J}_2$ to the following form
\begin{eqnarray}\label{eq:a1}
    \mathcal{J}_2 &=& -\frac{1}{\pi a_{\text{B}} \ell^2}\iint\limits_{\mathbb{R}^2} d^2 \boldsymbol{\rho} \exp\bigg( - \frac{\boldsymbol{\rho}^2}{2 \ell^2} \bigg) \ln \frac{z}{|\boldsymbol{\rho} + \boldsymbol{\rho}_0|} = \nonumber \\ & = &- \frac{2}{a_{\text{B}}}\bigg[ \ln \frac{z}{\ell} - \frac{1}{2} \ln 2 + \frac{1}{2} C - \nonumber \\ & & - \frac{1}{2\pi}\int\limits_{0}^{\infty} \tau d \tau \exp(-\tau^2) \mathcal{J}(\tau)\bigg],
\end{eqnarray}
where $\mathcal{C}$ is Euler's constant, numerically equal to $\mathcal{C} \simeq 0.57721$. By $\mathcal{J}(\tau)$ we mean the following integral
\begin{equation}\label{eq:a2}
    \mathcal{J}(\tau) = \int\limits_{0}^{2\pi} d \varphi \ln \bigg( 1 + \frac{2 \rho_0}{\sqrt{2} \tau \ell}\cos\varphi + \frac{\rho_0^2}{2 \tau^2 \ell^2} \bigg),
\end{equation}
which can be calculated by isolating the perfect square in the expression under the logarithm
\begin{eqnarray}\label{eq:a3}
    \mathcal{J}(\tau) &=& \int\limits_{0}^{2 \pi} d\varphi \ln\bigg( 1 + \frac{2 \rho_0}{\sqrt{2} \tau \ell} \cos2\varphi + \frac{\rho_0^2}{2 \tau^2 \ell^2} \bigg) = \nonumber\\
	& & = 4 \pi \ln \frac{|1 - a| + 1 + a}{2} = \nonumber \\ & & = \begin{cases}
		0, \text{ when } a < 1\\
		4\pi \ln a, \text{ when } a \geq 1
	\end{cases},
\end{eqnarray}
where the notation $a = \rho_0/\sqrt{2} \tau \ell$ was introduced. Thus, integration over $\tau$ in $\mathcal{J}_2$ terminates at the value $\tau = \rho_0/\sqrt{2} \ell$, that is
\begin{eqnarray}\label{eq:a4}
    \mathcal{J}_2 &=&  -\frac{2}{a_{\text{B}}}\bigg[ \ln \frac{z}{\ell} - \frac{1}{2} \ln 2 + \frac{1}{2}\mathcal{C} - \nonumber \\ & & -\frac{1}{2 \pi} \int\limits_{0}^{\rho_0^2/2 \ell^2} dt \exp(-t) \ln \frac{\rho_0^2}{2 t \ell^2}\bigg].
\end{eqnarray}

\section{Asymptotic behavior of function $\Lambda$}\label{app:b}
Function $\Lambda(x)$ has the form
\begin{equation}
    \Lambda(x) = \int\limits_{0}^x dy e^{-y} \ln \frac{x}{y}.
\end{equation}

For small $x \rightarrow 0$ we have
\begin{eqnarray}\label{b:limit_small_x}
    \Lambda(x) = x + \mathcal{O}(x^2), ~~~~~ x \rightarrow 0,
\end{eqnarray}
where the fact was used that $df(x)/dx = (1/x)\int_0^x dy e^{-y} \rightarrow 1$ for $x \rightarrow 0$. Taking the derivative in a similar way, in another limiting case $x \rightarrow +\infty$ we can obtain
\begin{eqnarray}\label{b:limit_big_x}
    \Lambda(x) = \ln x + \mathcal{C} + \mathcal{O}(e^{-x}), ~~~~~ x \rightarrow +\infty.
\end{eqnarray}

\nocite{*}

\bibliography{article}

\end{document}